\documentclass{article}

\usepackage{spconf,amsmath,graphicx}
\usepackage{calc}
\usepackage[acronym, toc, nonumberlist]{glossaries}
\usepackage{multirow}
\usepackage{pifont}
\usepackage{siunitx}
\usepackage{subcaption}
\usepackage{tikz}
\usepackage{xcolor}
\usepackage{xspace}
\usetikzlibrary{shapes,arrows}
\newacronym{AM}{AM}{acoustic model}
\newacronym{AMI}{AMI}{Augmented Multi-party Interaction}
\newacronym{ARQ}{ARQ}{automatic repeat request}
\newacronym{ASR}{ASR}{automatic speech recognition}
\newacronym[longplural={bi-directional long-short term memories}]{BLSTM}{BLSTM}{bi-directional long-short term memory}
\newacronym{BSS}{BSS}{blind speech separation}
\newacronym{CART}{CART}{classification and regression tree}
\newacronym{CE}{CE}{cross entropy}
\newacronym{CDp}{CDp}{context dependent phoneme}
\newacronym{CNN}{CNN}{convolutional neural network}
\newacronym{CPC}{CPC}{contrastive predictive coding}
\newacronym{CTC}{CTC}{connectionist temporal classification}
\newacronym{DCT}{DCT}{discrete cosine transform}
\newacronym{DL}{DL}{deep learning}
\newacronym{DNN}{DNN}{deep neural network}
\newacronym{DNN-HMM}{DNN-HMM}{deep neural network hidden Markov model}
\newacronym{ELBO}{ELBO}{evidence lower bound}
\newacronym{EM}{EM}{expectation maximization}
\newacronym{FE}{FE}{feature extractor}
\newacronym{FIR}{FIR}{finite impulse response}
\newacronym{FFNN}{FFNN}{feed-forward neural network}
\newacronym{G2P}{G2P}{grapheme-to-phoneme conversion}
\newacronym{GAN}{GAN}{generative adversarial network}
\newacronym{GMM}{GMM}{Gaussian mixture model}
\newacronym{GMM-HMM}{GMM-HMM}{Gaussian mixture model hidden Markov model}
\newacronym{HMM}{HMM}{hidden Markov model}
\newacronym{IHM}{IHM}{individual headset microphones}
\newacronym{IIR}{IIR}{infinite impulse response}
\newacronym{LAS}{LAS}{listen-attend-spell}
\newacronym{LDA}{LDA}{linear discriminant analysis}
\newacronym{LM}{LM}{language model}
\newacronym[longplural={long-short term memories}]{LSTM}{LSTM}{long-short term memory}
\newacronym{MDM}{MDM}{multiple distant microphones}
\newacronym{MFCC}{MFCC}{Mel-frequency cepstral coefficients}
\newacronym{MRES}{MRES}{multi-resolutional}
\newacronym{NLP}{NLP}{natural language processing}
\newacronym{NN}{NN}{neural network}
\newacronym{OOV}{OOV}{out-of-vocabulary}
\newacronym{PC2}{$\text{PC}^{\text{2}}$}{Paderborn Center for Parallel Computing}

\newacronym{RNA}{RNA}{recurrent neural aligner}
\newacronym{RNN}{RNN}{recurrent neural network}
\newacronym{RSAN}{RSAN}{recurrent selective attention network}
\newacronym{SAT}{SAT}{speaker adaptive training}
\newacronym{SDM}{SDM}{single distant microphone}
\newacronym{SDR}{SDR}{signal-to-distortion ratio}
\newacronym{SC}{SC}{supervised convolutional}
\newacronym{sMBR}{sMBR}{state-level minimum Bayes risk}
\newacronym{STFT}{STFT}{short time Fourier transform}
\newacronym{TDNN}{TDNN}{time delay neural network}
\newacronym[longplural={time-frequencies}]{tf}{tf}{time-frequency}
\newacronym{VAD}{VAD}{voice activity detection}
\newacronym{VTLN}{VTLN}{vocal tract length normalization}
\newacronym{cpWER}{cpWER}{concatenated minimum-permutation word error rate}
\newacronym{WER}{WER}{word error rate}
\newacronym{WP}{WP}{work package}
\newacronym{WPE}{WPE}{weighted prediction error}
\newacronym{WSJ}{WSJ}{Wall Street Journal}

\newcommand{\tab}{Table~}
\newcommand{\fig}{Figure~}
\newcommand{\sect}{Section~}
\newcommand{\ivec}{i-vector\xspace}
\newcommand{\ivecs}{i-vectors\xspace}
\newcommand{\relu}{ReLU\xspace}
\newcommand{\devclean}{\textit{dev-clean}\xspace}
\newcommand{\devother}{\textit{dev-other}\xspace}
\newcommand{\testclean}{\textit{test-clean}\xspace}
\newcommand{\testother}{\textit{test-other}\xspace}
\newcommand{\chime}[1]{CHiME~#1}
\newcommand{\cmark}{\ding{51}}
\newcommand{\xmark}{\ding{55}}

\newcommand{\captionTableFeatureCombination}{\Glspl{WER} [\%] for different feature combinations. The pre-trained \gls{AM} was used in all cases and the wav2vec model where applicable. Pre-training and supervised training are conducted on the 960h of LibriSpeech. All results are obtained on \devother using the 4-gram \gls{LM}.}
\newcommand{\captionTableBestResults}{\Glspl{WER} [\%] with a combination of Gammatone, wav2vec and \ivecs (comb.) on \testclean and \testother compared to other models from the literature. The pre-trained \gls{AM} (12.5 epochs) is used and we train further 8 epochs. Sequence discriminative training contributes another 1.5 epochs resulting in 22 supervised epochs in total. Trf denotes Transformer and Cnf-Trnsd stands for the Conformer Transducer from \cite{google2020conformer}. Pre-training and supervised training are conducted on the 960h of LibriSpeech.}
\newcommand{\captionTablePerplexities}{\Gls{LM} perplexities on LibriSpeech test sets \cite{irie2019trafolm}.}
\newcommand{\captionTableTraining}{Overview of the possibility to update the parameters during unsupervised and supervised training as well as the number of parameters for Gammatone (GT), supervised convolutional (SC) and wav2vec (w2v) features. The number of parameters in the \gls{AM} varies only because of the different input dimensions to the first \gls{BLSTM} layer.}
\newcommand{\captionTableUnsupervised}{\Glspl{WER} [\%] with and without unsupervised pre-training of wav2vec features. Pre-training and supervised training are conducted on the 960h of LibriSpeech and 16 epochs of supervised training were performed here. All results are obtained on \devother using the 4-gram \gls{LM}.}
\newcommand{\captionTablePretraining}{\Glspl{WER} [\%] with and without supervised pre-training of the \gls{AM}. In case of pre-training, the \gls{AM} was trained for 12.5 epochs with GT features and we train for 8 further epochs with the stated features (see \sect\ref{sec:hybrid_am}). Pre-training and supervised training are conducted on the 960h of LibriSpeech. All results are obtained on \devother using the 4-gram \gls{LM}.}

\usepackage[absolute,showboxes]{textpos}

\TPMargin{5pt}

\newcommand{\copyrightstatement}{
    \begin{textblock}{0.84}(0.08,0.92)
         \noindent
         \footnotesize
         Copyright 2021 IEEE.
         Published in the 2021 IEEE Automatic Speech Recognition and Understanding Workshop (ASRU) (ASRU 2021), scheduled for 14-18 December 2021 in Cartagena, Colombia.
         Personal use of this material is permitted.
         However, permission to reprint/republish this material for advertising or promotional purposes or for creating new collective works for resale or redistribution to servers or lists, or to reuse any copyrighted component of this work in other works, must be obtained from the IEEE.
         Contact: Manager, Copyrights and Permissions / IEEE Service Center / 445 Hoes Lane / P.O. Box 1331 / Piscataway, NJ 08855-1331, USA.
         Telephone: + Intl. 908-562-3966.
    \end{textblock}
}

\title{On Architectures and Training for\\Raw Waveform Feature Extraction in ASR}

\name{Peter Vieting$^1$, Christoph L\"uscher$^{1,2}$, Wilfried Michel$^{1,2}$, Ralf Schl\"uter$^{1,2}$, Hermann Ney$^{1,2}$}
\address{
  $^1$Human Language Technology and Pattern Recognition Group,\\ Computer Science Department, RWTH Aachen University, 52074 Aachen, Germany\\
  $^2$AppTek GmbH, 52062 Aachen, Germany}

\begin{document}

\copyrightstatement
\maketitle
\begin{abstract}
With the success of neural network based modeling in \gls{ASR}, many studies investigated acoustic modeling and learning of feature extractors directly based on the raw waveform.
Recently, one line of research has focused on unsupervised pre-training of feature extractors on audio-only data to improve downstream \gls{ASR} performance.
In this work, we investigate the usefulness of one of these front-end frameworks, namely wav2vec, in a setting without additional untranscribed data for hybrid \gls{ASR} systems.
We compare this framework both to the manually defined standard Gammatone feature set, as well as to features extracted as part of the acoustic model of an \gls{ASR} system trained supervised.
We study the benefits of using the pre-trained feature extractor and explore how to additionally exploit an existing acoustic model trained with different features.
Finally, we systematically examine combinations of the described features in order to further advance the performance.
\end{abstract}

\glsresetall

\section{Introduction}
Traditionally, \gls{ASR} systems rely on hand-crafted feature extraction methods such as log Mel filterbank, \gls{MFCC} or Gammatone features \cite{lat:icassp2016, zhou2020tedlium2, du2020ustc}.
The advance of neural modeling in ASR raised the question whether feature extraction should rather be a part of the neural \gls{AM}.
This could avoid potential information loss which might result in suboptimal performance.
Besides some early methods using \glspl{FFNN} \cite{tuske2014raw}, mainly convolutional approaches have been proposed.
They seem especially suitable for this task as they can operate similar to previous feature extraction methods but take advantage of learnable filters \cite{palaz2015convolutional, golik15:cnn, sainath2015learning, tuske2018:waveform}.
In \cite{ravanelli2018sincnet}, the authors propose to use parameterized sinc functions to implement band-pass filters with few parameters in a way that is easy to interpret.

Recently, one line of research on learnable features for \gls{ASR} focused on architectures which can be pre-trained on audio-only data in an unsupervised fashion \cite{facebook2019wav2vec, facebook2019vqwav2vec, facebook2020wav2vec2, facebook2020xlsr, deepmind2020cpc}.
Untranscribed audio data is easier to obtain and the pre-training enables exploiting data which could otherwise not be incorporated.
This is shown to be helpful in multiple cases, especially for low resource languages \cite{facebook2020xlsr, deepmind2020cpc}.
Yet, it remains unclear whether these approaches offer additional benefits for research tasks with a closed set of training data.
Additionally, none of these features has been tested with a hybrid \gls{NN}/\gls{HMM} system beyond basic \gls{CTC} to the best of our knowledge.

While lots of different architectures for \gls{ASR} systems have been emerging in recent years \cite{lat:icassp2016, graves2012rnnt, zhou2021phoneme-transducer, variani2020hat}, hybrid \gls{NN}/\gls{HMM} systems constitute the state-of-the-art on many tasks like TED-LIUM release 2 \cite{zhou2020tedlium2}, \chime{6} \cite{du2020ustc} or AMI \cite{kanda2019mdm} and achieve competitive results on LibriSpeech \cite{facebook2020hybrid}.
It is therefore worthwhile to investigate how they can be further improved and how they compare to other high performance systems.
This work aims at investigating and comparing the usefulness of different features -- including learnable ones -- for hybrid \gls{NN}/\gls{HMM} systems.
Besides using a pre-trained \gls{FE}, we explore how a strong existing \gls{AM} trained with different features on the same task can be exploited to reduce the training effort needed in order to deploy the new front-end.
In the same way, \ivecs are integrated to retrofit the system with speaker adaptation.

The feature extraction methods used in this work as well as some considerations regarding the hybrid \gls{ASR} systems are presented in \sect\ref{sec:methods}. The experiments are described in \sect\ref{sec:experiments} and discussed in \sect\ref{sec:results}. \sect\ref{sec:conclusions} concludes the paper.

\section{Methods} \label{sec:methods}

\begin{figure}
\centering
\begin{tikzpicture}[auto,>=latex']
    \tikzstyle{block} = [draw, shape=rectangle, minimum height=3em, minimum width=3em, node distance=2.1cm, line width=1pt, text width=1.5cm, align=center]
    \node (input) {$s_1^M$};
    \node [block, right of=input] (fe) {Feature Extractor};
    \node [node distance=0.5cm, right of=fe] (dummy) {};
    \node [block, right of=dummy] (am) {Acoustic Model};
    \node [node distance=2.3cm, right of=am] (output) {$p_t\!\left(\phi|\textbf{x}_1^T\right)$};
    \begin{scope}[line width=1pt]
         \draw[->] (input) -- (fe);
         \draw[->] (fe) -- node [above] {$\textbf{x}_1^T$} (am);
         \draw[->] (am) -- (output);
    \end{scope}
\end{tikzpicture}
\caption{Overall \gls{NN} architecture with waveform samples $s_1^M$ as input, the \gls{FE} output $\textbf{x}_1^T$ and the \gls{AM} which models the \gls{CDp} label posterior probability $p_t\!\left(\phi|\textbf{x}_1^T\right)$.}
\label{fig:fe_am}
\end{figure}
\begin{figure*}[htb]
\begin{minipage}[t]{.48\linewidth}
  \centering
  \includegraphics[width=.9\linewidth]{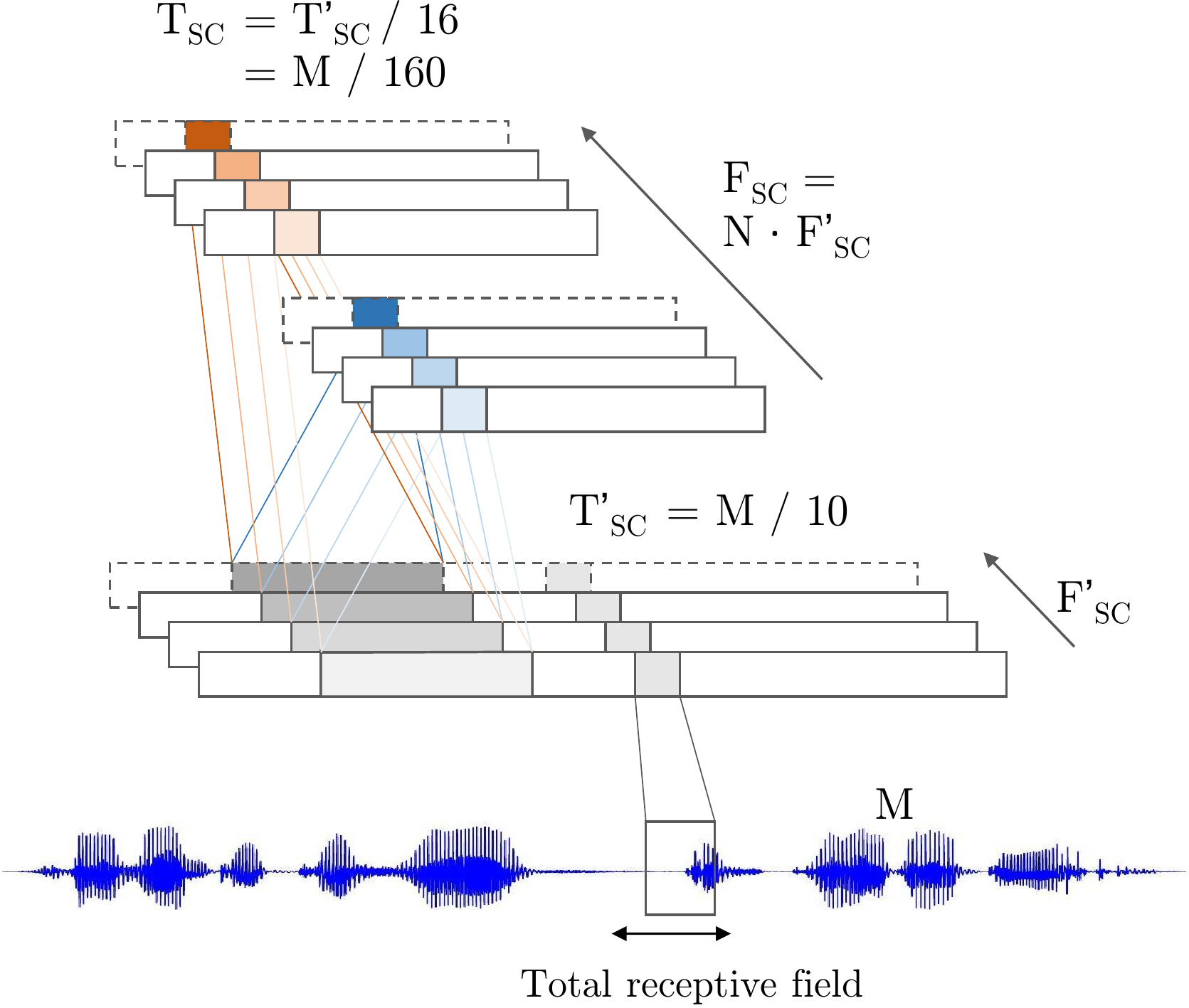}
  \centerline{(a) Supervised convolutional features}\medskip
\end{minipage}
\hfill
\begin{minipage}[t]{0.48\linewidth}
  \centering
  \includegraphics[width=.9\linewidth]{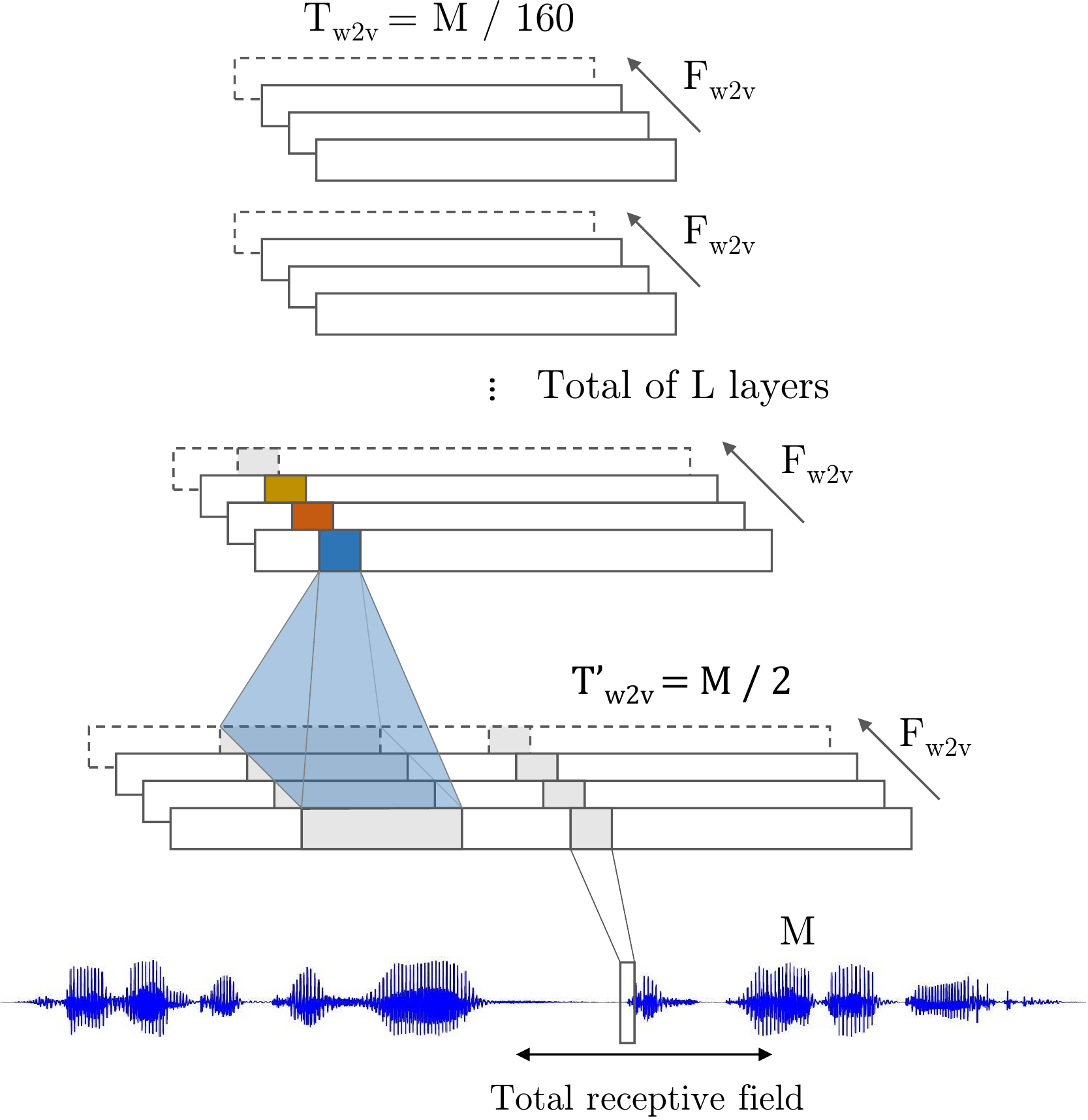}
  \centerline{(b) wav2vec}\medskip
\end{minipage}
\caption{
Illustration of the supervised convolutional and wav2vec architectures.
The horizontal rectangles represent feature channels over time.
The last dashed rectangle in each layer indicates that there are many channels per layer.
The connecting lines between layers show how an entry is calculated based on the previous layer's activations.
Typically, an entry is calculated based on all feature channels of the previous layer.
E.g., in wav2vec, the first filter is represented in blue and the entries are computed using all previous channels.
The other filters represented by other colors work analogously.
In the supervised convolutional features extraction, the second layer pools each channel over time with multiple low pass filters.
So the entries in the first channel are computed only based on values from the previous first channel.
$M$ is the number of waveform samples and the subsampling in time can be seen from $T_{SC}$ and $T_{w2v}$ respectively (ignoring boundary effects).
The final sequence length is the same, $T_{SC}=T_{w2v}$.
The number of features is $F_{w2v}$ and $F_{SC}$ respectively where $N$ is the number of low-pass filters for pooling in the second \gls{SC} layer.
While the \gls{SC} architecture contains 2 layers, wav2vec large comprises $L=19$ layers in total.
Activation functions and normalizations are not depicted here.
Note that the sizes are not drawn to scale.
}
\label{fig:mres_wav2vec}
\end{figure*}

\subsection{Feature Extraction}
In the following, the feature extraction methods used in this work are explained. 
Note that for traditional hand-crafted feature extraction methods, there is a clear border between the \gls{FE} and the subsequent \gls{AM}, as depicted in \fig\ref{fig:fe_am}.
If the neural network operates directly on the audio samples, this border blurs.
Nevertheless, for the sake of clarity, we refer to the layers that replace the hand-crafted feature extraction as \gls{FE} and to the remaining layers which are unchanged as \gls{AM}.

\subsubsection{Gammatone Features} \label{sec:gammatone}
Gammatone features are based on the Gammatone filter \cite{aertsen1980gammatone} which is designed to mimic the human auditory filter.
They were first introduced for large vocabulary \gls{ASR} in \cite{Schlueter2007}. 
After pre-emphasizing the speech signal, a filterbank of Gammatone filters with center frequencies sampled from the Greenwood function \cite{greenwood1990cochlear} is applied.
Temporal integration of the filter outputs' absolute values is typically performed using a Hanning window of \SI{25}{\milli\second} width with \SI{10}{\milli\second} shifts.
After $10^{th}$ root compression, a cepstral decorrelation using the \gls{DCT} and normalization techniques are applied.

\subsubsection{Supervised Convolutional Features}
The described Gammatone feature extraction pipeline motivated the approach in \cite{tuske2018:waveform}.
First, a convolutional layer for the purpose of \gls{tf} decomposition is applied on the raw waveform.
In contrast to other works which train acoustic models directly on the waveform, the following envelope extraction is not performed by non-parameterized function such as max pooling, but rather with a rectification followed by low-pass filters which are shared between \gls{tf} filters.
By introducing multiple ($N$) low-pass filters, a multi-resolutional processing can be achieved.
Finally, an additional non-linearity, e.g. logarithmic or root compression, may be applied, and the resulting features can be interpreted as critical band energies.
Since the convolutional layers of this method are trained supervised, we refer to the features as \gls{SC} features.
The architecture is shown in \fig\ref{fig:mres_wav2vec}~(a).

\subsubsection{wav2vec Features} \label{sec:wav2vec}
The wav2vec features were first introduced in \cite{facebook2019wav2vec} and aim at learning audio representations in an unsupervised fashion from unlabeled data.
These representations are supposed to improve downstream \gls{ASR} performance.
The network architecture consists of an encoder and a context network.
A regular model with 5 layers in the encoder and 9 layers in the context network as well as a large model with 7 and 12 layers respectively were presented.
All convolutional layers have 512 channels and use the \relu activation function.
In the encoder, skip connections are used to allow for better convergence of the large model.
The output of the context network can be used as input to the \gls{ASR} system.
\fig\ref{fig:mres_wav2vec}~(b) depicts the model.

A contrastive loss is used as the objective during unsupervised pre-training. As no pre-training is conducted in this work, we refrain from giving the details.
There have been follow up works using quantization and self-attention \cite{facebook2019vqwav2vec, facebook2020wav2vec2}, however, we focus on the fully convolutional version.

\subsubsection{\ivecs}
A popular speaker adaptation method in \gls{ASR} is the addition of \ivecs\cite{kitza19:interspeech, xiong2018microsoft, kanda2018hitachi} which aim at conveying speaker characteristics.
We follow the procedure described in \cite{kitza19:interspeech} to train and extract the \ivecs.
Specifically, the silence frames in the input waveform are filtered out first.
Then, Gammatone features with a context of 9 frames and a subsequent \gls{LDA} to reduce the feature dimension to 60 are used to estimate the \ivecs.
The dimensionality of the \ivecs themselves is set to 200 and they are normalized to have unit Euclidean norm.
We extract one \ivec per utterance.

\subsection{Hybrid ASR Systems}\label{sec:hybrid_am}
The \gls{ASR} experiments in this work are carried out using hybrid \gls{NN}/\gls{HMM} systems.
We take the model described in \cite{luescher2019librispeech} as the baseline for our work and use the same \glspl{CDp} and the same alignments.
The \gls{AM} architecture consists of 6 \gls{BLSTM}\glsunset{LSTM} layers with 1000 units for each direction followed by a linear layer with softmax activation which maps to the 12001 \gls{CDp} labels clustered by a \gls{CART} \cite{luescher2019librispeech}.

In \cite{luescher2019librispeech}, all \glspl{AM} trained with the frame-wise \gls{CE} loss were trained from scratch, i.e., the network parameters were initialized randomly at the beginning of the training.
However, if a pre-trained model is already available, its parameters may be used in the same way a pre-trained \gls{FE} is used.
This can reduce the training effort needed for deploying a new set of features.

A possible example of how to exploit this is presented in \cite{zhang2020learning}, where the goal is to obtain noise invariant features. 
A model trained on clean data is virtually divided into \gls{FE} and classifier by grouping lower and upper layers.
Then, the model is further trained on noisy data while keeping the classifier constant or updating its parameters with a small learning rate.
The learning rate for the \gls{FE} is not reduced and its parameters are either initialized randomly or from the clean model.

In this work, we use the best model obtained by frame-wise \gls{CE} training in \cite{luescher2019librispeech} as a baseline to use the parameters of the \gls{BLSTM} layers as well as the softmax output layer and continue training the model after resetting the learning rate.
The baseline model was trained with 50-dimensional Gammatone features.
If the dimension of the features differs from this, the first layer's parameters can not be used because the weight matrix has a different shape.
We circumvent this by initializing the first layer randomly and only adopt the parameters of the subsequent layers.
Optionally, a random initialization of further subsequent layers may be applied.
Unlike \cite{zhang2020learning}, we update all parameters with the same learning rate.

While most experiments are done using the frame-wise \gls{CE} loss, we also deploy a lattice-based version of the \gls{sMBR} criterion to conduct sequence discriminative training.
The lattices are created using the frame-wise \gls{CE}-trained system.
Then, we reset the learning rate and continue training now using the \gls{sMBR} criterion and an additional \gls{CE} smoothing.
Note that this learning rate reset and continued training resembles the procedure described above.
However, none of the \gls{BLSTM} layers is initialized randomly.

\subsection{Language Models}
During recognition, the official 4-gram \gls{LM} as well as a custom \gls{LSTM} \gls{LM} are used in the first pass decoding \cite{beck2019:lstmlm1pass}.
Additionally, the lattices are rescored utilizing a Transformer \gls{LM} \cite{irie2019trafolm}.
The \gls{LSTM} \gls{LM} consists of a linear layer followed by two \gls{LSTM} layers while the Transformer \gls{LM} is built by 96 layers with 8 attention heads and both have an output softmax layer over the full 200k vocabulary.
Both are identical with the ones used in \cite{luescher2019librispeech}.
The perplexities can be found in \tab\ref{tab:perplexities}.
\begin{table}[h]
\centering
\caption{\captionTablePerplexities}
\label{tab:perplexities}
\begin{tabular}{|c|r|r|r|r|}
\hline
   \gls{LM} &     \multicolumn{2}{c|}{dev} &    \multicolumn{2}{c|}{test} \\\cline{2-5}
            & {clean} & {other} & {clean} & {other} \\\hline\hline
     4-gram &   151.7 &   140.6 &   158.1 &   145.7 \\\cline{1-5}
 \gls{LSTM} &    60.2 &    60.2 &    64.8 &    61.7 \\\cline{1-5}
Transformer &    53.2 &    54.2 &    57.6 &    55.0 \\
\hline
\end{tabular}
\end{table}

\section{Experiments} \label{sec:experiments}
The experiments carried out for this work are presented in the following section.
As in \cite{luescher2019librispeech}, the RWTH open-source toolkits RASR \cite{rybach2011rasr, wiesler2014rasr} and RETURNN\footnote{Training configuration files for the \gls{AM} will be made available at \texttt{https://github.com/rwth-i6/returnn-experiments/}.} \cite{longdoetsch2016returnn, zeyer2018returnn} were used for training and recognition.

\subsection{LibriSpeech Dataset}
The experiments are carried out on LibriSpeech which contains English read speech sampled at \SI{16}{\kilo\hertz}.
Both the supervised \gls{ASR} training as well as the unsupervised pre-training in \cite{facebook2019wav2vec} are done on the 960 hours training data, so no external data was used for the wav2vec features.
The \glspl{LM} are trained on the transcriptions of the 960 hours as well as the additional 800M-word text-only data.
Testing is done on the common \devclean, \devother, \testclean and \testother subsets.

\subsection{Features}
All features are extracted from the raw waveform inside \mbox{RETURNN}.
This allows for an easy integrability but also makes the feature extraction a fully differentiable part of the neural network which can be helpful in case the \gls{ASR} system is integrated with neural pre-processing steps such as speech separation or enhancement.

For the Gammatone features, we therefore apply a convolutional layer with 50 Gammatone filters and a size of 640 samples which corresponds to \SI{40}{\milli\second}.
This is in contrast to the Gammatone feature extraction in \cite{luescher2019librispeech}, where RASR is used to extract Gammatone features and an implementation using \gls{IIR} filters is deployed.
The remaining steps are carried out as described in \sect\ref{sec:gammatone}.
Finally, we deploy batch normalization.

For the supervised convolutional features, we apply 150 filters for the \gls{tf} decomposition each with a size of 256 samples and a stride of 10 and take the absolute values of the outputs.
The envelope extraction is performed by 5 filters with a size of 40 samples and a stride of 16.
After taking the root of the absolute values, a layer normalization completes the feature extraction.

The wav2vec features are extracted as described in \cite{facebook2019wav2vec}.
We adapt the architecture denoted as \textit{Large} in the original work, i.e., a convolutional encoder network consisting of 7 layers and a convolutional context network of 12 layers are used.
The resulting features are 512-dimensional.
Parameters of pre-trained models were made available by the authors and we use them in our experiments.
As stated before, the pre-training was conducted on the LibriSpeech training data and no additional untranscribed audio data was used.
The unsupervised pre-training is not subject of this work, however, we continue training the parameters using the supervised loss during \gls{ASR} training.

\begin{table}[t]
\centering
\caption{\captionTableTraining}
\label{tab:training}
\begin{tabular}{|c|c|c|c|c|}
\hline
Features & \multicolumn{2}{c|}{Trainable} &  \multicolumn{2}{c|}{\#params} \\\cline{2-5}
         &    unsup. &      sup. & \gls{FE} & \gls{AM} \\\hline\hline
      GT &    \xmark &    \xmark &      35k &     152M \\\cline{1-1}\cline{3-5}
      SC &           &    \cmark &      40k &     158M \\\cline{1-2}\cline{4-5}
     w2v &    \cmark &           &      29M &     156M \\
\hline
\end{tabular}
\end{table}
\tab\ref{tab:training} summarizes these features and outlines that while the Gammatone features are always extracted with a fixed set of parameters, the parameters of the supervised convolutional features can be updated during the supervised training and the wav2vec model can additionally even be pre-trained in an unsupervised fashion.
Additionally, we can observe that the number of parameters for Gammatone and supervised convolutional features is comparable, while the wav2vec architecture exceeds this by a large factor.
Note that the number of parameters for the Gammatone feature extraction is based on the \gls{FIR} implementation described above.

The subsampling in the wav2vec encoder is chosen such that a feature frame is obtained every \SI{10}{\milli\second} which corresponds to the frame rate typically used for Gammatone features, e.g. in \cite{luescher2019librispeech}.
This is convenient for training the hybrid \gls{ASR} system with the frame-wise \gls{CE} criterion as the alignments obtained using Gammatone features can still be used.
Additionally, it facilitates frame-wise feature combination.
One frame of the wav2vec features in the regular model has a total receptive field of about \SI{210}{\milli\second} whereas it increases to about \SI{810}{\milli\second} for the large model.
This is significantly larger than for the convolutional implementation of the Gammatone features or the features of the supervised convolutional approach, where the total receptive field corresponds to about \SI{65}{\milli\second} or \SI{40}{\milli\second} of audio, respectively.

When comparing different features, it seems natural to apply feature combination for possible further improvements.
This is done using a frame-wise concatenation of all features here.
The \ivecs are computed per utterance, therefore, this single vector is repeated and concatenated in each frame.

\section{Experimental Results} \label{sec:results}

\subsection{Effect of Training Strategies}
\begin{table}[t]
\centering
\caption{\captionTableUnsupervised}
\label{tab:unsupervised}
\begin{tabular}{|c|c|S|}
\hline
Features & Epochs unsup. & {WER} \\\hline\hline
     w2v &             0 &  10.1 \\\cline{2-3}
         &           129 &   9.7 \\
\hline
\end{tabular}
\end{table}
The effect of performing unsupervised pre-training of the wav2vec features is shown in \tab\ref{tab:unsupervised}.
The experiment with 0 epochs of unsupervised pre-training corresponds to the case where the wav2vec model is initialized randomly and only updated in supervised training.
We can observe that it helps to use pre-training even in a setting where no additional untranscribed data is used.
However, the relative gains are smaller than the ones on \gls{WSJ} in \cite{facebook2019wav2vec} when using additional data which is intuitive.

In the baseline model of \cite{luescher2019librispeech}, the pre-extracted input sequence of Gammatone features is split into chunks using a windowing process during mini-batch construction to speed up the training.
The window comprises 50 frames and is shifted by 25 frames, which corresponds to \SI{0.5}{\second} and \SI{0.25}{\second} of the input signal, respectively.
A collection of chunks obtained by this windowing process then makes up one mini-batch.
Now, since the network input is the raw waveform, we perform chunking on the audio directly. 
As described in \sect\ref{sec:wav2vec}, the total receptive field of the wav2vec large model is about \SI{810}{\milli\second}, which exceeds the chunk size of \SI{0.5}{\second}.
Therefore, chunking was performed with a size of \SI{3}{\second} and a shift of \SI{1.5}{\second} for our experiments in \tab\ref{tab:unsupervised}.
We observed, however, that reducing it to a size of \SI{1}{\second} and a shift of \SI{0.5}{\second} reduced the training time by about 25\% while sacrificing only 0.1\% absolute in \gls{WER} which is why we apply this setting for the subsequent experiments.

\begin{table}[t]
\centering
\caption{\captionTablePretraining}
\label{tab:pretraining}
\begin{tabular}{|c|c|c|c|S|}
\hline
                         Features & Pretrained GT-AM & \multicolumn{2}{c|}{Epochs} & {WER} \\\cline{3-4}
                                  &    (12.5 epochs) & unsup. &   sup. &       \\\hline\hline
GT \cite{luescher2019librispeech} &               no &      - &   12.5 &   9.6 \\\cline{1-2}\cline{4-5}
                               GT &              yes &        &      8 &   8.7 \\\cline{1-2}\cline{4-5}
                               SC &               no &        &     16 &   9.6 \\\cline{2-2}\cline{4-5}
                                  &              yes &        &      8 &   9.0 \\\cline{1-5}
                              w2v &               no &    129 &     16 &   9.7 \\\cline{2-2}\cline{4-5}
                                  &              yes &        &      8 &   8.8 \\
\hline
\end{tabular}
\end{table}
\tab\ref{tab:pretraining} depicts the effect of using a pre-trained \gls{AM}.
We use the baseline from \cite{luescher2019librispeech} which was trained for 12.5 epochs and train for 8 additional epochs here.
Consistent improvements across all features can be seen.
The parameters of the pre-trained \gls{AM} are exploited as described in \sect\ref{sec:hybrid_am} and only the first \gls{BLSTM} layer is initialized randomly.
Random initialization of further \gls{BLSTM} layers resulted in inferior performance.

\subsection{Comparison of Architectures}
As observed in \tab\ref{tab:pretraining}, the performance of the \gls{ASR} system using wav2vec features is on par with the baseline in \cite{luescher2019librispeech} when training the \gls{AM} from scratch as well as with a pre-trained \gls{AM}.
This proves the features' suitability for hybrid systems.
Similarly, the supervised convolutional features' performance is in the same ballpark, although slightly worse when using the pre-trained \gls{AM}.
The gap compared to the Gammatones matches the degradation of 3-4\% relative found in \cite{tuske2018:waveform} for \gls{FFNN}.
However, the results using an \gls{LSTM} back-end were significantly worse in \cite{tuske2018:waveform} which we do not observe here.
A possible explanation is that the front-end parameters were trained with a \gls{FFNN} and kept fixed during the \gls{LSTM} training in \cite{tuske2018:waveform} while we train them together with the \gls{BLSTM} layers.
To be fair, it should be noted that the wav2vec features might benefit from a system combination effect as parameters trained by the authors of \cite{facebook2019wav2vec} are used in our experiments.

\subsection{Feature Combination}
\begin{table}[t]
\centering
\caption{\captionTableFeatureCombination}
\label{tab:feature_combination}
\begin{tabular}{|>{\centering}p{\widthof{i-vec}}|>{\centering}p{\widthof{i-vec}}|>{\centering}p{\widthof{i-vec}}|>{\centering}p{\widthof{i-vec}}|S|}
\hline
 \multicolumn{4}{|c|}{Features} & {WER} \\\cline{1-4}
       GT &       w2v &        SC &     i-vec &       \\\hline\hline
$\bullet$ &         ~ &         ~ &         ~ &   8.7 \\\cline{1-5}
        ~ & $\bullet$ &         ~ &         ~ &   8.8 \\\cline{1-5}
        ~ &         ~ & $\bullet$ &         ~ &   9.0 \\\cline{1-5}
$\bullet$ & $\bullet$ &         ~ &         ~ &   8.4 \\\cline{1-5}
$\bullet$ &         ~ & $\bullet$ &         ~ &   8.8 \\\cline{1-5}
        ~ & $\bullet$ & $\bullet$ &         ~ &   8.7 \\\cline{1-5}
$\bullet$ &         ~ &         ~ & $\bullet$ &   8.2 \\\cline{1-5}
        ~ & $\bullet$ &         ~ & $\bullet$ &   8.1 \\\cline{1-5}
        ~ &         ~ & $\bullet$ & $\bullet$ &   8.4 \\\cline{1-5}
$\bullet$ & $\bullet$ &         ~ & $\bullet$ &   8.0 \\\cline{1-5}
$\bullet$ & $\bullet$ & $\bullet$ & $\bullet$ &   8.1 \\
\hline
\end{tabular}
\end{table}
Next, \tab\ref{tab:feature_combination} shows an overview of the results obtained by different features and their combinations.
Combining Gammatone and wav2vec features improves the performance while adding supervised convolutional features to the other ones shows only marginal differences.
A significant gain of 5-8\% relative is attained when adding \ivecs to each of the previous features.
This demonstrates the feasibility of retrofitting an \gls{AM} with speaker adaptation by integrating \ivecs in the presented way.
It also shows that even for the learnable and pre-trained wav2vec features, additional speaker information is helpful.
The best performance can be achieved with a combination of Gammatone, wav2vec and \ivecs.
A further iteration of resetting the learning rate and initializing the first layer randomly did not yield an improvement for this model.

\begin{table}[t]
\centering
\caption{\captionTableBestResults}
\label{tab:best_results}
\setlength\tabcolsep{0.5pt}
\begin{tabular}{|c|c|c|c|c|c|S|S|}
\hline
                         Features &           AM & \multicolumn{2}{c|}{Epochs} &   Seq. &   \gls{LM} &   \multicolumn{2}{c|}{{WER}} \\\cline{3-4}\cline{7-8}
                                  &              & Unsup. &   Sup. &  dscr. &            & {clean} & {other} \\\hline\hline
GT \cite{luescher2019librispeech} & \gls{BLSTM}- &      - &   12.5 &     no &     4-gram &     4.4 &    10.0 \\\cline{1-1}\cline{3-4}\cline{7-8}
                            Comb. &       Hybrid &    129 &   20.5 &        &            &     3.9 &     8.6 \\\cline{1-1}\cline{3-5}\cline{7-8}
GT \cite{luescher2019librispeech} &              &      - &     14 &    yes &            &     3.8 &     8.8 \\\cline{1-1}\cline{3-4}\cline{7-8}
                            Comb. &              &    129 &     22 &        &            &     3.5 &     8.1 \\\cline{1-1}\cline{3-4}\cline{6-8}
GT \cite{luescher2019librispeech} &              &      - &     14 &        & \gls{LSTM} &     2.6 &     5.5 \\\cline{1-1}\cline{3-4}\cline{7-8}
                            Comb. &              &    129 &     22 &        &            &     2.5 &     5.2 \\\cline{1-1}\cline{3-4}\cline{6-8}
GT \cite{luescher2019librispeech} &              &      - &     14 &        &       +Trf &     2.3 &     5.0 \\\cline{1-1}\cline{3-4}\cline{7-8}
                            Comb. &              &    129 &     22 &        &            &     2.2 &     4.7 \\\cline{1-8}
  LgMel \cite{facebook2020hybrid} &   Trf-Hybrid &      - &    200 &     no &    4gr+Trf &     2.1 &     4.2 \\\cline{1-4}\cline{6-8}
 LgMel \cite{google2020conformer} &    Cnf-Trnsd &     {} &     {} &        &       LSTM &     1.9 &     3.9 \\
\hline
\end{tabular}
\end{table}
We perform sequence discriminative training for 1.5 epochs and recognition with the \gls{LSTM} \gls{LM} as well as lattice rescoring using the Transformer \gls{LM} for the best model.
Additionally, an optimization of decoding parameters is conducted for the sequence discriminatively trained models.
The results are depicted in \tab\ref{tab:best_results}.
The comparison shows that the improvements over \cite{luescher2019librispeech} that we achieve for the frame-wise \gls{CE} trained model are stronger on the \textit{other} set than on the \textit{clean} one.
Furthermore, the benefit of sequence discriminative training is smaller than it was in \cite{luescher2019librispeech}.
We suspect that while the positive influence of the new training criterion persists, the effect of a learning rate reset and some further epochs of training is already included in the frame-wise \gls{CE} trained model here, unlike in \cite{luescher2019librispeech}, which could explain the observation.

In contrast, the relative improvements obtained by using the \gls{LSTM} \gls{LM} remain almost constant.
As observed for the frame-wise \gls{CE}-trained model, the gains over our previous best system are stronger on the \textit{other} set, where we achieve a relative improvement of \SI{6}{\percent}, while the result is \SI{4}{\percent} relative better on \testclean.

The results presented in \cite{facebook2020hybrid} are the best ones currently published on LibriSpeech with a hybrid system.
The main reasons for the remaining gap in performance of our model are likely the different layer architecture as well as the time spent for supervised training of the \gls{AM}.
While 200 epochs were carried out in \cite{facebook2020hybrid}, we achieve our final results after training the \gls{AM} for only 22 epochs.
A further gap can be observed when comparing to the state-of-the-art performance reported in \cite{google2020conformer} where the number of training epochs conducted is unclear.

\section{Conclusion} \label{sec:conclusions}
We apply learnable feature extractors, namely wav2vec and supervised convolutional features, in a hybrid \gls{BLSTM}/\gls{HMM} \gls{ASR} system and show their suitability for this type of model.
Using a pre-trained wav2vec module helps, however, it does not outperform the non-trainable features and features trained only with a supervised loss in a setting without additional untranscribed data.
It is demonstrated that an existing \gls{AM} trained on Gammatone features can be exploited in a similar way for all features.
A combination of Gammatone and wav2vec features as well as \ivecs leads to the best results outperforming our previous best system by \SI{4}{\percent} and \SI{6}{\percent} relative on \testclean and \testother, respectively.

\section{Acknowledgements}
This project has received funding from the European Research Council (ERC) under the European Union's Horizon 2020 research and innovation programme (grant agreement n\textsuperscript{o}~694537, project "SEQCLAS").
The work reflects only the authors' views and the European Research Council Executive Agency (ERCEA) is not responsible for any use that may be made of the information it contains.

This work was partially supported by the project HYKIST funded by the German Federal Ministry of Health on the basis of a decision of the German Federal Parliament (Bundestag).

We thank Markus Kitza for extracting the \ivecs as well as Wei Zhou for his help on the \gls{LSTM} \gls{LM} decoding.
We also thank the authors of wav2vec \cite{facebook2019wav2vec} for providing their model parameters.

\bibliographystyle{IEEEbib}
\bibliography{mybib}

\end{document}